\documentclass[twocolumn,showpacs,preprintnumbers,amsmath,amssymb,aps,prl,superscriptaddress]{revtex4-1}
\usepackage{color}
\usepackage{graphicx}
\usepackage{textcomp} 
\usepackage[caption=false]{subfig} 
\begin{document}
\title{Relaxation and Intermediate Asymptotics of a Rectangular Trench in a Viscous Film}
\author{Oliver B\"{a}umchen}\thanks{These authors contributed equally to this work.}
\affiliation{Department of Physics \& Astronomy and the Brockhouse Institute for Materials Research, McMaster University, Hamilton, Canada}
\author{Michael Benzaquen}\thanks{These authors contributed equally to this work.}
\affiliation{Laboratoire de Physico-Chimie Th\'eorique, UMR CNRS Gulliver 7083, ESPCI, Paris, France}
\author{Thomas Salez}
\affiliation{Laboratoire de Physico-Chimie Th\'eorique, UMR CNRS Gulliver 7083, ESPCI, Paris, France}
\author{Joshua D.\ McGraw}\thanks{Present address: Department of Experimental Physics, Saarland University, D-66041 Saarbr\"ucken, Germany}
\affiliation{Department of Physics \& Astronomy and the Brockhouse Institute for Materials Research, McMaster University, Hamilton, Canada}
\author{Matilda Backholm}
\affiliation{Department of Physics \& Astronomy and the Brockhouse Institute for Materials Research, McMaster University, Hamilton, Canada}
\author{Paul Fowler}
\affiliation{Department of Physics \& Astronomy and the Brockhouse Institute for Materials Research, McMaster University, Hamilton, Canada}
\author{Elie Rapha\"{e}l}
\affiliation{Laboratoire de Physico-Chimie Th\'eorique, UMR CNRS Gulliver 7083, ESPCI, Paris, France}
\author{Kari Dalnoki-Veress}\email{dalnoki@mcmaster.ca}
\affiliation{Department of Physics \& Astronomy and the Brockhouse Institute for Materials Research, McMaster University, Hamilton, Canada}
\affiliation{Laboratoire de Physico-Chimie Th\'eorique, UMR CNRS Gulliver 7083, ESPCI, Paris, France}
\date{\today}
\begin{abstract}
The surface of a thin liquid film with nonconstant curvature flattens as a result of capillary forces. While this leveling is driven by \emph{local} curvature gradients, the \emph{global} boundary conditions greatly influence the dynamics. Here, we study the evolution of rectangular trenches in a polystyrene nanofilm. Initially, when the two sides of a trench are well separated, the asymmetric boundary condition given by the step height controls the dynamics. In this case, the evolution results from the leveling of two noninteracting steps. As the steps broaden further and start to interact, the global symmetric boundary condition alters the leveling dynamics. We report on full agreement between theory and experiments for: the capillary-driven flow and resulting time dependent height profiles; a crossover in the power-law dependence of the viscous energy dissipation as a function of time as the trench evolution transitions from two noninteracting to interacting steps;  and the convergence of the profiles to a universal self-similar attractor that is given by the Green's function of the linear operator describing the dimensionless linearized thin film equation.
\end{abstract}
\maketitle
Thin films are prevalent in applications such as lubricants, coatings for optical and electronic devices, and nanolithography to name just a few. However, it is also known that the mobility of polymers can be altered in thin films~\cite{shin07NMT, bodiguel06PRL,fakhraai08SCI, si05PRL,oconnell2005}. Thus, understanding the dynamics of thin films in their liquid state is essential to gaining control of pattern formation and relaxation on the nanoscale \cite{leveder08APL,teisseire11APL}. For example, high-density data storage in thin polymer films is possible by locally modifying a surface with a large 2-D array of atomic force microscope probes~\cite{vettinger02}. This application relies on control of the time scales of the flow created by a surface profile to produce or erase a given surface pattern. In contrast to this technologically spectacular example is the surface of freshly applied paint which relies on the dynamics of leveling to provide a lustrous surface. 

Much has been learned about the physics of thin films from dewetting, where an initially flat film exposes the substrate surface to reduce the free energy of the system~\cite{Srolovitz, brochard90CJP,reiter92PRL, seemann01PRL, fetzer05PRL, reiter05NM, vilmin06EPJE, baumchen09PRL, snoeijer10PRE, baumchen2012JCP}. Other approaches, for example studying the evolution of surface profiles originating from capillary waves~\cite{Tsui10}, embedding of nanoparticles~\cite{fakhraai08SCI}, or those created by an external electric field~\cite{barbero09PRL, closa} have also been utilized to explore mobility in thin films. Although the capillary-driven leveling of a nonflat surface topography in a thin liquid film has been reported in various studies~\cite{rognin11PRE, rognin12JVS}, experiments with high resolution compared to a general theory that relates time scales to spatial geometries and properties of the liquid are still lacking. Recently, we studied the leveling of a \textit{stepped film}: a new nanofluidic tool to study the properties of polymers in thin film geometries~\cite{jdmlev1, McGraw12, salez12a, Salez12b, McGraw13}.
\begin{figure}  
\includegraphics[width=1.0\columnwidth]{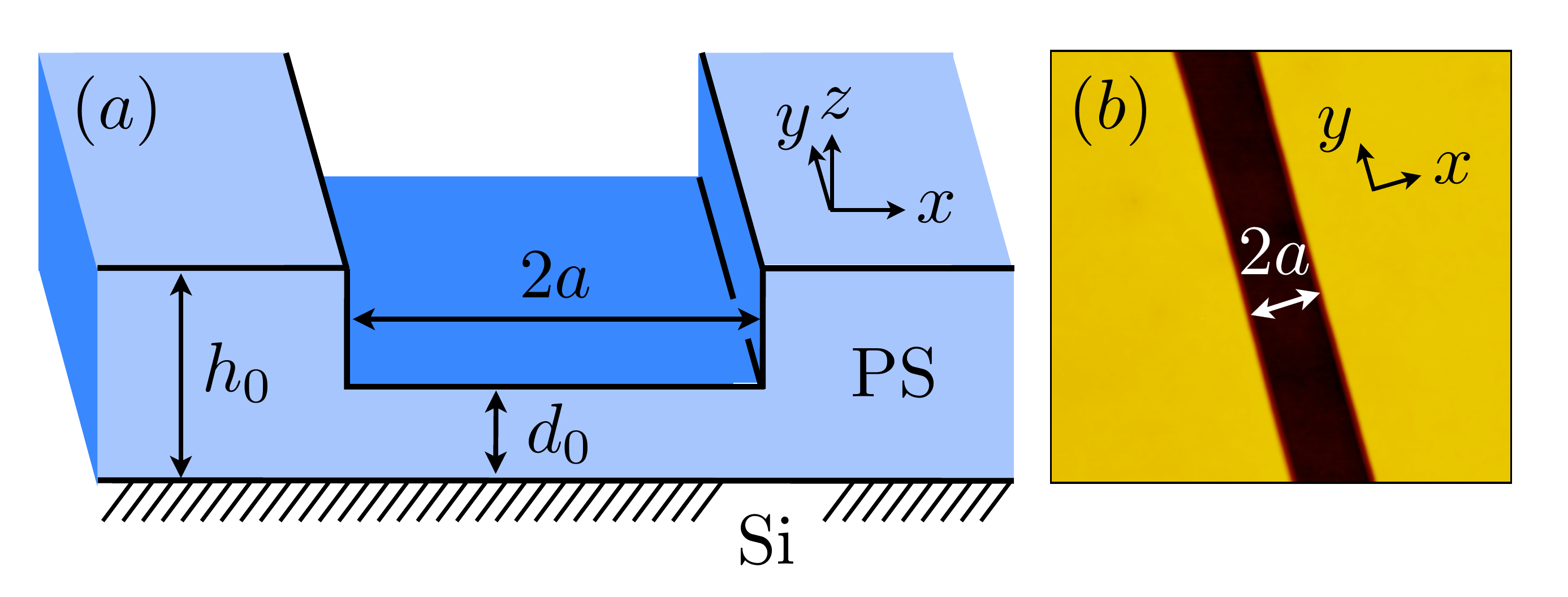}  
\caption{(a) Schematic of the initial geometry of a  polystyrene film on a silicon wafer (not to scale: $2a$ is a few $\mu$m's, while $h_0$ and $d_0 \sim 100 \,$nm). (b) Atomic force microscopy image of a typical trench (image width $30\,\mu$m, height range 115 nm). }
\label{schematic}
\end{figure} 
\begin{figure*}[t!]
\begin{center}     
\includegraphics[width=2.05\columnwidth]{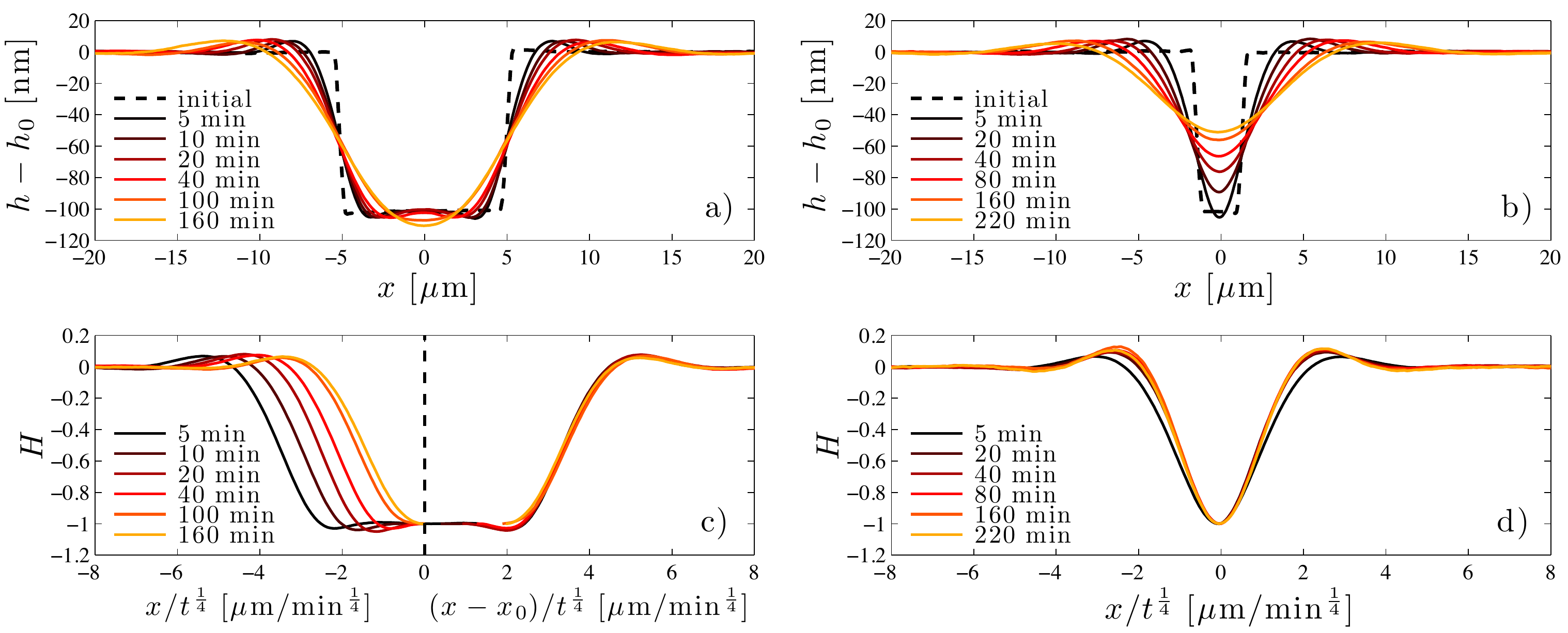}  
\end{center}
\caption{(top row) Experimental surface profiles at various times ($5 \leq t \leq 220\,$min). The initial rectangular trenches are shown with dashed lines and  exhibit widths of (a) $a$ = 5.3\,$\mu$m and (b) $a$ = 1.6 $\mu$m. (bottom row) The normalized height data of (a) and (b) are rescaled in accordance with the scaling predictions of the theory. The right panel of (c) shows the data shifted to illustrate the collapse of the self-similar profiles, where $x_0(t)$ is a horizontal shift for each right front. Profiles in (a) and (c) show the approach of two noninteracting steps, whereas in (b) and (d) the steps are interacting and the trench depth decreases.}
\label{profiles}
\end{figure*} 

In this work, we study the leveling of perfect \textit{trenches} in thin liquid polymer films as shown in Fig.\,\ref{schematic}. A trench is composed of two opposing steps separated by a distance $2a$. This structure evolves because of gradients in the Laplace pressure resulting from gradients in the curvature. Initially, the trench levels as two noninteracting steps, with an asymmetric global boundary condition set by the height $h_0$ of the film at the top of the step, and the height $d_0$ of the film at the bottom of the step. However, as the steps broaden they interact with one another,  which results in an overall symmetric boundary condition, with $h_0$ on both sides. As will be shown, the crossover from asymmetric to symmetric boundary conditions modifies the scaling law for the energy dissipation in the film. This striking feature reveals the fundamental role played by the boundary conditions in the evolution of global quantities of the system. Furthermore, the trench geometry provides an ideal illustration of the theory of \textit{intermediate asymptotics}~\cite{baren03TXT}: a key tool underlying historical examples such as the Reynolds drag force~\cite{Reynolds}, Kelvin's nuclear explosion~\cite{Taylor1, Taylor2}, and scaling theory in general, in a situation that is highly relevant for industry and fundamental nanorheology. Intermediate asymptotics theory is based on self-similarity, addresses complex nonlinear partial differential equations, and provides solutions at intermediate times --  a time range far enough from the initial state so that details of the initial condition can be forgotten, and far enough from the trivial equilibrium state. The value of those intermediate solutions is thus to bring a certain generality by offering an alternative to the principle of superposition, which is lacking in nonlinear physics. In the work presented here, the profile does not evolve self-similarly for early times. However, after some transient period it becomes self-similar and converges to a universal attractor that depends on the boundary condition: the intermediate asymptotic solution~\cite{Benzaquen13}.  

Polymer films exhibiting rectangular trench geometries, as illustrated in Fig.\,\ref{schematic}, were prepared as follows. Polystyrene (PS) films were spincast from a toluene (Fisher Scientific, Optima grade) solution onto  freshly cleaved mica sheets (Ted Pella Inc.). The PS has molecular weight $M_\mathrm{w}=31.8\,$~kg/mol with polydispersity index of 1.06 (Polymer Source, Inc.). The samples were annealed in a vacuum oven ($\sim \,10^{-5}$ mbar) for 24\,hrs at 130$\,^{\circ}$C. This temperature is well above the glass transition temperature ($T_{\mathrm{g}} \approx  100\, ^\circ \mathrm{C}$) and ensures removal of residual solvent and relaxation of the polymer chains. Films were then floated onto the surface of ultrapure water (18.2\,M$\Omega$cm, Pall, Cascada LS) and picked up with $1 \mathrm{\ cm} \times 1 \mathrm{\ cm}$ Si wafers (University Wafer). Si wafers were cleaned by exposure to air plasma (low power, Harrick Plasma, for 30 s), with subsequent rinses in ultrapure water, methanol (Fisher Scientific, Optima grade) and toluene. After the film was transferred to the Si, the floating transfer was repeated resulting in a second film with the same thickness on top of the first polymer film. The polymer films are not highly entangled and easily fracture on the surface of the water resulting in straight float gaps. These float gaps can be hundreds of $\mu$m long and only a few $\mu$m wide. Thus, the second transfer creates a rectangular trench such that $h_0=2d_0$, see Fig.\,\ref{schematic}. The edges of the trench were checked with atomic force microscopy (AFM, Veeco Caliber) and optical microscopy to ensure that there were no defects in the trench and the vicinity. We stress that the samples are prepared at room temperature, well below $T_{\mathrm{g}}$, and only flow when heated above $T_{\mathrm{g}}$.

In all cases studied, the width of the trench was constant and much smaller than its length. The problem can thus be safely treated as invariant in the $y$-coordinate along the length of the trench. Prior to each measurement, the initial condition was recorded using AFM, as shown in Figs.\,\ref{profiles}(a) and \ref{profiles}(b): the width of the trench $2a$, as well as the depth $h_0-d_0$ was determined from an analysis of height profiles. Independently, thicknesses were measured with AFM from float gaps or a small scratch made through the film to the substrate. Samples were annealed in ambient conditions~\cite{note1} at 140$\,^{\circ}$C on a hot stage (Linkam) with a heating rate of $ 90\,^{\circ}$C/min. Height profiles $z=h(x,t)$ representing the vertical distance between the substrate-liquid and liquid-air interfaces at position $x$ of the trenches were recorded using AFM after various times $t$, following a quench to room temperature. Figures\,\ref{profiles}(a) and \ref{profiles}(b) show the evolution of rectangular trenches for $h_0=2\,d_0=206$~nm, with $a=5.3\,\mu$m and $a=1.6\,\mu$m respectively, as a function of the annealing time $t$. We deliberately used two different initial widths in order to address the two different temporal regimes in a single rescaled description, as detailed below. As can be seen in Fig.\,\ref{profiles}(a), the profiles broaden and are accompanied by a bump at the top of the step and a dip at the bottom of the step that are characteristic of isolated steps~\cite{McGraw12}.  Initially, both sides of the trench in Fig.\,\ref{profiles}(a) level independently of each other until $t \sim 100\,$min, when both dips merge into a single minimum. The subsequent stage is characterized by the decreasing depth of the profile as shown in Fig.\,\ref{profiles}(b) while the profile continues to broaden. 

The relaxation process of the surface can be described by considering capillary driving forces originating from the nonconstant curvature of the surface~\cite{degennes03TXT}: Gradients in the Laplace pressure provide a driving force for the leveling of the surface topography. The Laplace pressure is given by $p(x,t) \approx -\gamma \partial_x^2 h$, where $\gamma$ is the air-liquid surface tension. The height scales in this study were chosen to be small enough that gravitational forces can be neglected~\cite{refBcite}, but sufficiently large to neglect disjoining forces~\cite{seeman2001PRL}. Moreover, we can safely exclude phenomena related to the polymer chain size, \emph{e.g.}\ confinement effects, as the film thicknesses are much larger than characteristic polymer chain length scales. The timescale of the longest relaxation time of PS (31.8\,kg/mol) at 140$\,^{\circ}$C is orders of magnitude shorter than the time scales considered here~\cite{bach03MAC}, thus we can treat the film as a simple Newtonian liquid. The theoretical description of the problem is thus based on the Stokes equation for highly viscous flows, and the lubrication approximation which states that all vertical length scales are small compared to horizontal ones~\cite{stillwagon88JAP, oron97RMP,craster09RMP,blossey}. A no-slip boundary condition at the liquid-substrate interface and the no-stress boundary condition at the liquid-air interface result in the familiar Poiseuille flow along the $z$ axis. Invoking conservation of volume leads to the capillary-driven thin film equation (TFE):
\begin{eqnarray}
\partial_t h + \frac{\gamma}{3\eta}\,\partial_x\left( h^3\partial_x^{\,3} h  \right)&=&0\,, \label{TFE}
\end{eqnarray}
where $\eta$ is the viscosity of the film~\cite{stillwagon88JAP, oron97RMP,craster09RMP}. The ratio $\gamma/\eta$ provides the typical speed of leveling and is termed the \textit{capillary velocity}. Equation~\eqref{TFE} is highly nonlinear and has no known general analytical solution. Nevertheless, it can be solved numerically using a finite difference algorithm~\cite{bertozziAMS98,salez12a}. Equation~\eqref{TFE} can also be linearized and solved analytically in the particular situation where the trench is a perturbation: $h_0\approx d_0$~\cite{Salez12b,Benzaquen13}. 
In this linear case, we recently showed that the solutions converge to a self-similar attractor~\cite{Benzaquen13}:
\begin{equation}
\label{conv}
h(x,t)\displaystyle\,\xrightarrow[t\rightarrow\infty]\,h_0+2\,(d_0-h_0)\,\, \mathcal{G}(U,T)\ ,
\end{equation}
where we introduced the two dimensionless variables:
\begin{eqnarray}
U&=&\frac{x}{t^{1/4}}\left(\frac{3\eta}{\gamma h_0^{\,3}}\right)^{1/4}\\ 
T&=&\frac{\gamma h_0^{\,3} t}{3\eta a^4}\ ,
\end{eqnarray}
as well as the Green's function of the linear operator describing the dimensionless linearized thin film equation:
\begin{equation}
\mathcal{G}(U,T)=\frac{1}{2\pi\, T^{1/4}}\int_{-\infty}^{\infty}dQ\,e^{-Q^4}e^{iQU}\ .
\end{equation}
Thus, the experimental profiles should collapse at long times when the vertical axis is normalised by the depth of the trench, $H=[h(x,t)-h_0]/[h_0-h(0,t)]$, and the horizontal axis is rescaled as $x/t^{1/4}$. We checked numerically that this statement is still true in the nonlinear case of Eq.~\eqref{TFE}~\cite{Benzaquen13}. The self-similar regime represents the intermediate asymptotics solution of this thin film problem~\cite{baren03TXT}. 

\begin{figure}[t]
\begin{center}     
\includegraphics[width=1\columnwidth]{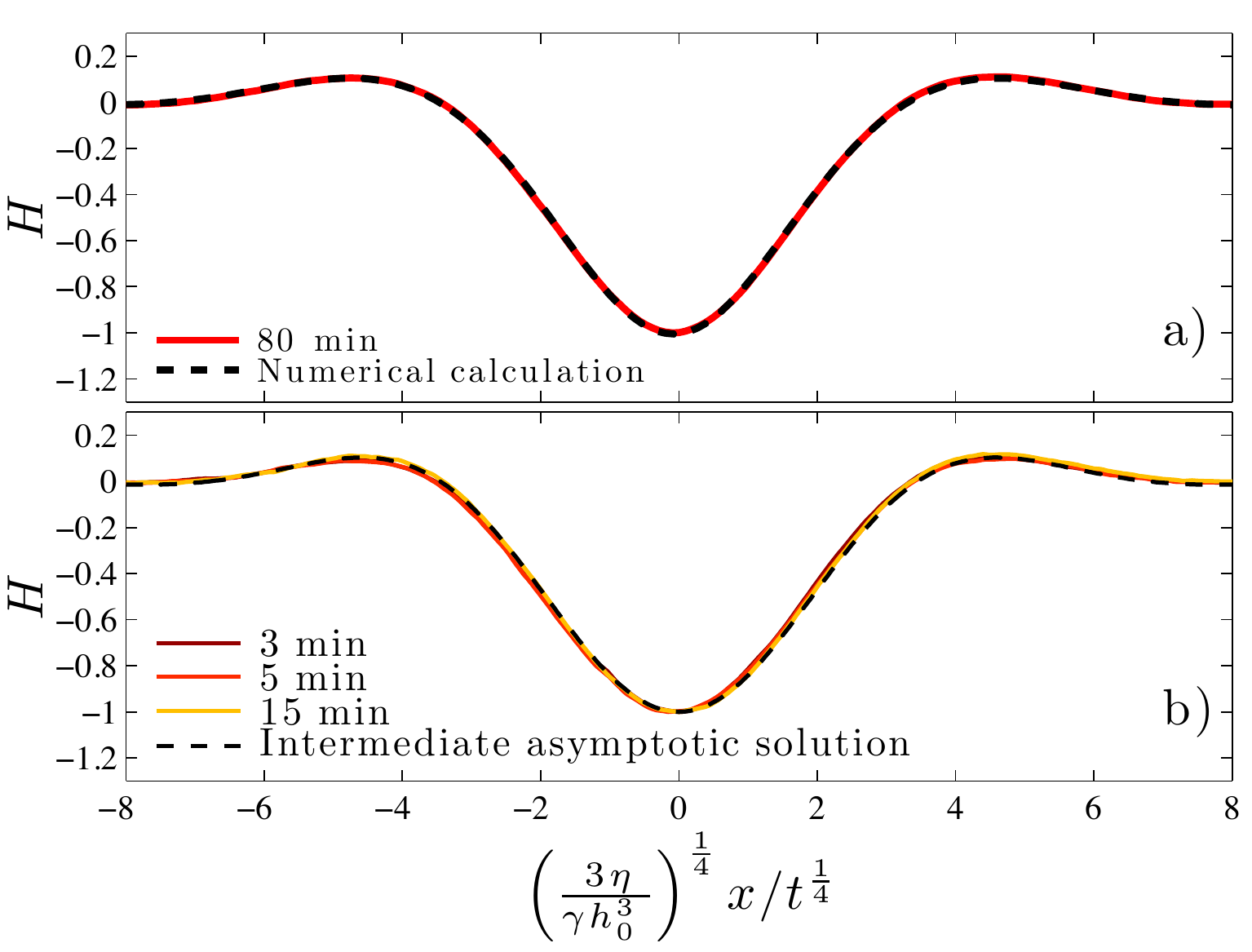}
\end{center}
\caption{Normalized height profiles for experiments (solid line) annealed at 140$\,^{\circ}$C and best-fit theory (dashed lines). (a) A trench with $a$ = 1.6\,$\mu$m, $h_0$ = 206\,nm, $d_0$ = 103\,nm after annealing for $t = 80$\,min, with corresponding numerical solution to Eq.~\ref{TFE}. (b) A trench where the depth is a small perturbation ($a$ = 2.5\,$\mu$m, $h_0$ = 611\,nm, $d_0$ = 59\,nm) at 140$\,^{\circ}$C. The dashed line is the analytical Green's function of the linear operator describing the linearised thin film equation.}
\label{fits}
\end{figure} 
We first consider the nonlinear situation, where $h_0=2\,d_0$. In Figs.\,\ref{profiles}(c) and \ref{profiles}(d) we show the rescaled data corresponding to the noninteracting (a) and interacting (b) regimes of the trench evolution. Excellent agreement with the theoretical predictions is obtained. The noninteracting steps all have the same profile in their frame of reference. This is demonstrated by shifting the data as shown in the right side of Fig.\,\ref{profiles}(c), where $x_0(t)$ is a horizontal shift for each right front. At long times, self-similarity in $x/t^{1/4}$ is obtained for the interacting steps. Thus, there are two distinct regimes: first, an initial stage where the steps broaden, and are self-similar in their frame of reference, but are not yet interacting; second, a crossover to a final self-similar stage where the steps have merged and the depth of the profile diminishes. These two distinct stages correspond to the crossover of the system from the asymmetric boundary condition of the noninteracting steps to the symmetric global boundary condition of the profile. As discussed, it is possible to numerically solve Eq.~\eqref{TFE} for the trench geometry~\cite{bertozziAMS98,salez12a}. The experimental data are in excellent agreement with nonlinear calculations as shown in Fig.\,\ref{fits}(a) for $t=80$~min. We note that this agreement is typical and obtained for all such comparisons. In the self-similar representation of the data, only a lateral stretch is required to match the experimental data and the calculation. According to Eq.~\eqref{TFE}, the stretching factor is directly related to the capillary velocity and we obtain $\eta/\gamma = 0.034\,$min/$\mu$m for PS (31.8\,kg/mol) at 140$\,^{\circ}$C. This value is in excellent agreement with literature values~\cite{wuPpoly, McGraw13} and consistent with the capillary velocity that can be determined from the noninteracting steps according to the technique described in~\cite{McGraw12}.

\begin{figure}[t!]
\includegraphics[width=1.01\columnwidth]{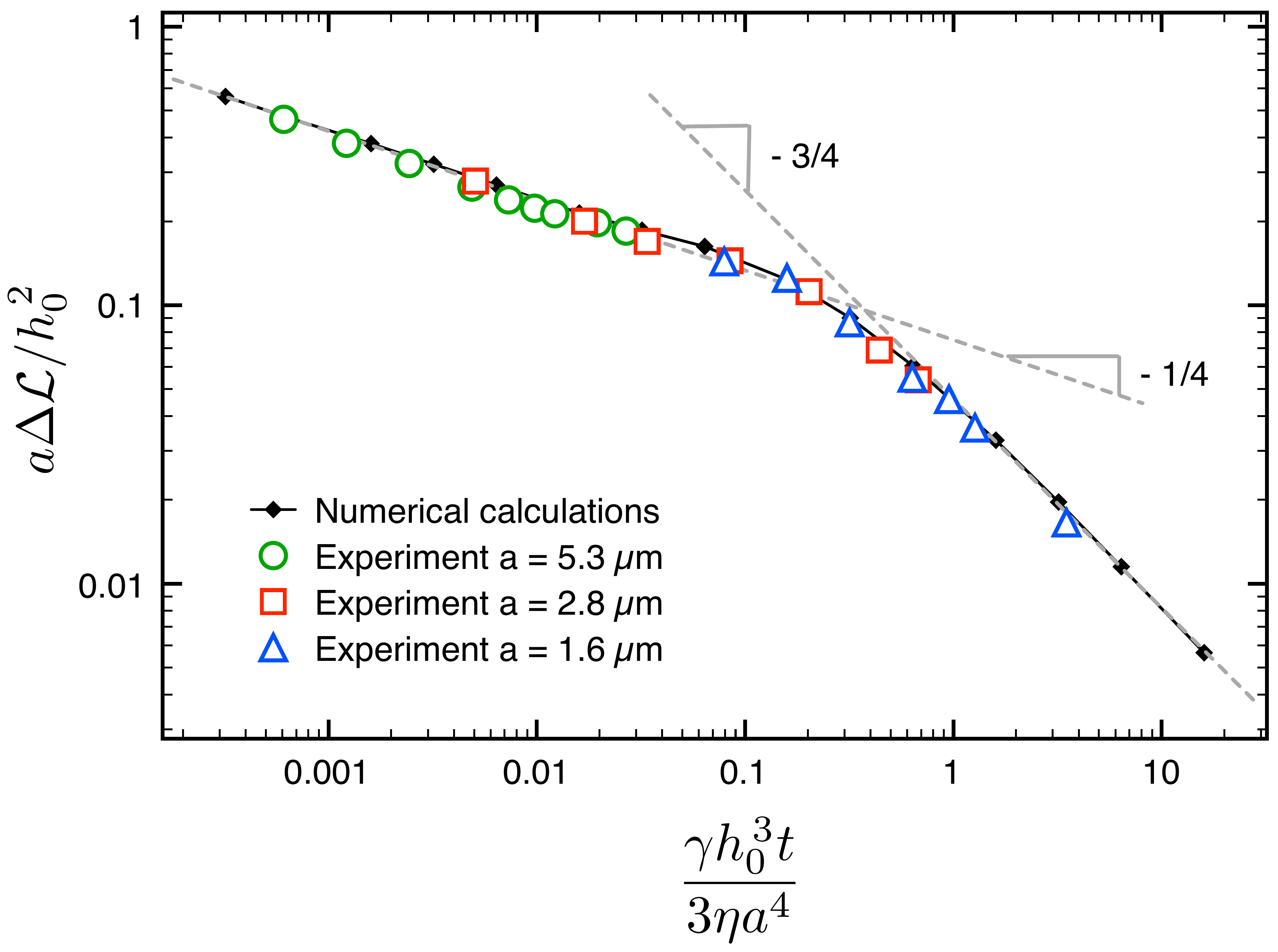}  
\caption{Excess contour length $\Delta \mathcal L$ of rectangular trenches (half-width $a$, reference film height $h_0$) in nondimensionalized representation as a function of the nondimensionalized time. Experimental data (open circles, rectangles and triangles) and nonlinear numerical calculations (solid circles) both show a transition from a $-1/4$ to a $-3/4$ power law scaling. Experimental error bars are less than $2\%$ of reported values.}
\label{energy}
\end{figure} 
With the theoretical tools described above, we are able to analyze the \textit{entire} evolution from noninteracting to interacting steps. A relevant quantity that can be extracted from the data is the excess contour length $\Delta \mathcal L$ as a function  of time. For small slopes, one has:
\begin{eqnarray}
\label{defen}
\Delta \mathcal L&\approx & \int dx \,\frac12\,(\partial_x h)^2\,.
\end{eqnarray}
From the specific dimensional invariance of Eq. \eqref{TFE}, one can show that:
\begin{eqnarray}
\frac{a  \Delta \mathcal L}{h_0^{\,2}}&=& \left(\frac{\gamma h_0^{\,3}t}{3\eta a^4} \right)^{-1/4}  f\left(  \frac{\gamma h_0^{\,3}t}{3\eta a^4}  ,\frac{h_0-d_0}{h_0} \right)\ , 
\end{eqnarray}
where $f$ is a function of two variables. This intrinsic similarity comes from the fact that any initial  profile $h_\lambda(\lambda x,0)=h(x,0)$ obtained from a horizontal stretch of factor $\lambda$ yields the evolution $h_\lambda (\lambda x, \lambda^4 t)=h(x,t)$ through Eq.~\eqref{TFE}. Thus, the excess contour length evolution of trenches with the same vertical aspect ratio, $h_0/d_0=2$, but with different widths can be rescaled in a single master plot as shown in Fig.\,\ref{energy}. In accordance with the theory, our experiments demonstrate a $\Delta \mathcal L \sim t^{-1/4}$ scaling regime for early times which corresponds to two independent relaxing steps~\cite{McGraw12}. Afterwards, the two steps interfere and there is a crossover to a long-time $\Delta \mathcal L \sim t^{-3/4}$ scaling regime. This observation can be understood by combining Eqs.~(\ref{conv}) and (\ref{defen}). Having done so we obtain:
\begin{equation}
\frac{a  \Delta \mathcal L}{h_0^{\,2}}\,\,\displaystyle\xrightarrow[t\rightarrow\infty]\,\,\frac{\Gamma(3/4)}{2^{7/4}\,\pi}\,\left(\frac{\gamma h_0^{\,3}t}{3\eta a^4}\right)^{-3/4}\left(\frac{h_0-d_0}{h_0}\right)^2\ ,
\end{equation}
which is valid in the linear case. The numerical calculations shown in Fig.\,\ref{energy} confirm this asymptotic $\Delta \mathcal L \sim t^{-3/4}$ scaling in the nonlinear case as well. This is expected since any profile will eventually become a perturbation at long time. The comparison of the data for three trenches with different widths and the numerical calculations shown in Fig.\,\ref{energy} clearly validates the theoretical predictions.

Finally, in order to study the intermediate asymptotics in more detail, we realized experimentally the linear case of a small surface perturbation: $h_0 \approx d_0$. This geometry evolves faster towards the intermediate asymptotic regime and a full analytic solution was obtained for the linearized thin film equation~\cite{Benzaquen13}. The sample was prepared by spincoating a thick PS film directly onto a clean Si wafer, while another, much thinner, PS film was prepared on a freshly cleaved mica sheet. The sample annealing, and floating of the top layer to prepare the trenches were done exactly as described for the previous $h_0=2\,d_0$ samples. In Fig.\,\ref{fits}(b), we show the self-similar profiles for three different annealing times, as well as the fit of the universal self-similar attractor given by the Green's function of the linearized thin film equation~\cite{Benzaquen13}. Excellent agreement is found with the only free parameter being the ratio: $\eta/\gamma= 0.016\,$min/$\mu$m. This value is again in very good agreement with expected literature values~\cite{wuPpoly, McGraw13} demonstrating the validity of the intermediate asymptotics.

To conclude, we provided new insights in the relaxation of a trench at the free surface of a viscous film. The surface relaxes due to the nonconstant curvature of the free interfaces which drives flow. We found that the transition from the asymmetric boundary condition of two noninteracting steps, to the symmetric boundary condition of the two interacting steps greatly influences the scaling of global properties such as the energy dissipation in the film. Specifically, the excess contour length crosses over from a $-1/4$ power law to a $-3/4$ power law with time in complete accordance with theoretical expectations. The profiles were found to be in quantitative agreement with full numerical resolution of the nonlinear thin film equation. Moreover, we have verifed experimentally that a small perturbative trench atop a flat film converges to a universal self-similar attractor that is given by the Green's function of the linear operator describing the linearized thin film equation. Therefore, while relevant to nanolithography, relaxation of a painted surface, and many other industrial processes that do not necessary involve polymers, the simple trench geometry developed here also provides an ideal application of intermediate asymptotics. For flat symmetric boundary condition, the properly rescaled self-similar attractor is \textit{universal} and should not depend on the initial profile, or on the viscous material used. 

The authors thank NSERC of Canada, the German Research Foundation (DFG) under grant numbers BA3406/2 and SFB 1027, the \'Ecole Normale Sup\'{e}rieure of Paris, the Fondation Langlois and the ESPCI Joliot Chair for financial support.


\begin{thebibliography}{99}
\expandafter\ifx\csname natexlab\endcsname\relax\def\natexlab#1{#1}\fi
\expandafter\ifx\csname bibnamefont\endcsname\relax
\def\bibnamefont#1{#1}\fi
\expandafter\ifx\csname bibfnamefont\endcsname\relax
\def\bibfnamefont#1{#1}\fi
\expandafter\ifx\csname citenamefont\endcsname\relax
\def\citenamefont#1{#1}\fi
\expandafter\ifx\csname url\endcsname\relax
\def\url#1{\texttt{#1}}\fi
\expandafter\ifx\csname urlprefix\endcsname\relax\def\urlprefix{URL }\fi
\providecommand{\bibinfo}[2]{#2}
\providecommand{\eprint}[2][]{\url{#2}}



\bibitem[{\citenamefont{O'Connell et~al.}(2005)\citenamefont{O'Connell, and McKenna}}]{oconnell2005}
\bibinfo{author}{\bibfnamefont{P. A.}~\bibnamefont{O'Connell}}, \bibnamefont{and}
\bibinfo{author}{\bibfnamefont{G. B.}~\bibnamefont{McKenna}},
\bibinfo{journal}{Science} \textbf{\bibinfo{volume}{307}},
\bibinfo{pages}{1760} (\bibinfo{year}{2005}).
  
\bibitem[{\citenamefont{Si et~al.}(2005)\citenamefont{Si, Massa, Dalnoki-Veress, Brown, and Jones}}]{si05PRL}
\bibinfo{author}{\bibfnamefont{L.}~\bibnamefont{Si}},
\bibinfo{author}{\bibfnamefont{M. V.}~\bibnamefont{Massa}},
\bibinfo{author}{\bibfnamefont{K.}~\bibnamefont{Dalnoki-Veress}},
\bibinfo{author}{\bibfnamefont{H. R.}~\bibnamefont{Brown}}, \bibnamefont{and}
\bibinfo{author}{\bibfnamefont{R. A. L.}~\bibnamefont{Jones}},
\bibinfo{journal}{Phys. Rev. Lett.} \textbf{\bibinfo{volume}{94}},
\bibinfo{pages}{127801} (\bibinfo{year}{2005}).
  
\bibitem[{\citenamefont{Bodiguel and Fretigny}(2006)}]{bodiguel06PRL}
\bibinfo{author}{\bibfnamefont{H.}~\bibnamefont{Bodiguel}} \bibnamefont{and}
\bibinfo{author}{\bibfnamefont{C.}~\bibnamefont{Fretigny}},
\bibinfo{journal}{Phys. Rev. Lett.} \textbf{\bibinfo{volume}{97}},
\bibinfo{pages}{266105} (\bibinfo{year}{2006}).
  
\bibitem[{\citenamefont{Shin et~al.}(2007)\citenamefont{Shin, Obukhov, Chen, Huh, Hwang, Mok, Dobriyal, Thiyagarajan, and Russell}}]{shin07NMT}
\bibinfo{author}{\bibfnamefont{K.}~\bibnamefont{Shin}},
\bibinfo{author}{\bibfnamefont{S.}~\bibnamefont{Obukhov}},
\bibinfo{author}{\bibfnamefont{J.-T.} \bibnamefont{Chen}},
\bibinfo{author}{\bibfnamefont{J.}~\bibnamefont{Huh}},
\bibinfo{author}{\bibfnamefont{Y.}~\bibnamefont{Hwang}},
\bibinfo{author}{\bibfnamefont{S.}~\bibnamefont{Mok}},
\bibinfo{author}{\bibfnamefont{P.}~\bibnamefont{Dobriyal}},
\bibinfo{author}{\bibfnamefont{P.}~\bibnamefont{Thiyagarajan}},
\bibnamefont{and} \bibinfo{author}{\bibfnamefont{T.}~\bibnamefont{Russell}},
\bibinfo{journal}{Nat Mater.} \textbf{\bibinfo{volume}{6}},  
\bibinfo{pages}{961} (\bibinfo{year}{2007}).

\bibitem[{\citenamefont{Fakhraai and Forrest}(2008)}]{fakhraai08SCI}
\bibinfo{author}{\bibfnamefont{Z.}~\bibnamefont{Fakhraai}} \bibnamefont{and}
\bibinfo{author}{\bibfnamefont{J. A.}~\bibnamefont{Forrest}},
\bibinfo{journal}{Science} \textbf{\bibinfo{volume}{319}},
\bibinfo{pages}{600} (\bibinfo{year}{2008}).
 
\bibitem[{\citenamefont{Leveder et~al.}(2008)\citenamefont{Leveder, Landis, and Davoust}}]{leveder08APL}
\bibinfo{author}{\bibfnamefont{T.}~\bibnamefont{Leveder}},
\bibinfo{author}{\bibfnamefont{S.}~\bibnamefont{Landis}}, \bibnamefont{and}
\bibinfo{author}{\bibfnamefont{L.}~\bibnamefont{Davoust}},
\bibinfo{journal}{Appl. Phys. Lett.} \textbf{\bibinfo{volume}{92}},
\bibinfo{pages}{013107} (\bibinfo{year}{2008}).

\bibitem[{\citenamefont{Teisseire et~al.}(2011)\citenamefont{Teisseire, Revaux, Foresti, and Barthel}}]{teisseire11APL}
\bibinfo{author}{\bibfnamefont{J.}~\bibnamefont{Teisseire}},
\bibinfo{author}{\bibfnamefont{A.}~\bibnamefont{Revaux}},
\bibinfo{author}{\bibfnamefont{M.}~\bibnamefont{Foresti}}, \bibnamefont{and}
\bibinfo{author}{\bibfnamefont{E.}~\bibnamefont{Barthel}},
\bibinfo{journal}{Appl. Phys. Lett.} \textbf{\bibinfo{volume}{98}},
\bibinfo{pages}{013106} (\bibinfo{year}{2011}).

\bibitem[{\citenamefont{Vettiger et~al.}(2002)\citenamefont{Vettinger, Cross, Despont, Drechsler, D\"urig, Gotsmann, H\"aberle, Lantz, Rothuizen, Stutz, Binnig}}]{vettinger02}
\bibinfo{author}{\bibfnamefont{P.}~\bibnamefont{Vettiger}},
\bibinfo{author}{\bibfnamefont{G.}~\bibnamefont{Cross}},
\bibinfo{author}{\bibfnamefont{M.} \bibnamefont{Despont}},
\bibinfo{author}{\bibfnamefont{U.}~\bibnamefont{Drechsler}},
\bibinfo{author}{\bibfnamefont{U.}~\bibnamefont{D\"urig}},
\bibinfo{author}{\bibfnamefont{B.}~\bibnamefont{Gotsmann}},
\bibinfo{author}{\bibfnamefont{W.}~\bibnamefont{H\"aberle}},
\bibinfo{author}{\bibfnamefont{M. A.}~\bibnamefont{Lantz}},
\bibinfo{author}{\bibfnamefont{H. E.}~\bibnamefont{Rothuizen}},
\bibinfo{author}{\bibfnamefont{R.}~\bibnamefont{Stutz}},
\bibnamefont{and} \bibinfo{author}{\bibfnamefont{G. K.}~\bibnamefont{Binnig}},
\bibinfo{journal}{IEEE Transactions on Nanotechnology} \textbf{\bibinfo{volume}{1}},
\bibinfo{pages}{39} (\bibinfo{year}{2002}).
  
\bibitem{Srolovitz} D. J. Srolovitz and S. A. Safran, J. Appl. Phys., \textbf{60}, 255 (1986).
  
\bibitem[{\citenamefont{Brochard~Wyart and Daillant}(1990)}]{brochard90CJP}
\bibinfo{author}{\bibfnamefont{F.}~\bibnamefont{Brochard~Wyart}}
\bibnamefont{and} \bibinfo{author}{\bibfnamefont{J.}~\bibnamefont{Daillant}},
\bibinfo{journal}{Can. J. of Phys.} \textbf{\bibinfo{volume}{68}},
\bibinfo{pages}{1084} (\bibinfo{year}{1990}).
  
\bibitem[{\citenamefont{Reiter}(1992)}]{reiter92PRL}
\bibinfo{author}{\bibfnamefont{G.}~\bibnamefont{Reiter}},
\bibinfo{journal}{Phys. Rev. Lett.} \textbf{\bibinfo{volume}{68}},
\bibinfo{pages}{75} (\bibinfo{year}{1992}).
  
\bibitem[{\citenamefont{Seemann et~al.}(2001{\natexlab{a}})\citenamefont{Seemann, Herminghaus, and Jacobs}}]{seemann01PRL}
\bibinfo{author}{\bibfnamefont{R.}~\bibnamefont{Seemann}},
\bibinfo{author}{\bibfnamefont{S.}~\bibnamefont{Herminghaus}},
\bibnamefont{and} \bibinfo{author}{\bibfnamefont{K.}~\bibnamefont{Jacobs}},
\bibinfo{journal}{Phys. Rev. Lett.} \textbf{\bibinfo{volume}{87}},
\bibinfo{pages}{196101} (\bibinfo{year}{2001}{\natexlab{a}}).
  
\bibitem[{\citenamefont{Fetzer et~al.}(2005)\citenamefont{Fetzer, Jacons, M\"unch, Wagner and Witelski}}]{fetzer05PRL}
\bibinfo{author}{\bibfnamefont{R.}~\bibnamefont{Fetzer}},
\bibinfo{author}{\bibfnamefont{K.}~\bibnamefont{Jacobs}}, 
\bibinfo{author}{\bibfnamefont{A.}~\bibnamefont{M\"unch}},
\bibinfo{author}{\bibfnamefont{B. A.}~\bibnamefont{Wagner}}, 
\bibnamefont{and}
\bibinfo{author}{\bibfnamefont{T. P.}~\bibnamefont{Witelski}},
\bibinfo{journal}{Phys. Rev. Lett.} \textbf{\bibinfo{volume}{95}},
\bibinfo{pages}{127801} (\bibinfo{year}{2005}).
  
\bibitem[{\citenamefont{Reiter et~al.}(2005)\citenamefont{Reiter, Hamieh, Damman, Sclavons, Gabriele, Vilmin, and Rapha\"{e}l}}]{reiter05NM}
\bibinfo{author}{\bibfnamefont{G.}~\bibnamefont{Reiter}},
\bibinfo{author}{\bibfnamefont{M.}~\bibnamefont{Hamieh}},
\bibinfo{author}{\bibfnamefont{P.}~\bibnamefont{Damman}},
\bibinfo{author}{\bibfnamefont{S.}~\bibnamefont{Sclavons}},
\bibinfo{author}{\bibfnamefont{S.}~\bibnamefont{Gabriele}},
\bibinfo{author}{\bibfnamefont{T.}~\bibnamefont{Vilmin}}, \bibnamefont{and}
\bibinfo{author}{\bibfnamefont{E.}~\bibnamefont{Rapha\"{e}l}},
\bibinfo{journal}{Nat. Mater.}
\textbf{\bibinfo{volume}{4}}, \bibinfo{pages}{754} (\bibinfo{year}{2005}).

\bibitem[{\citenamefont{Vilmin and Rapha\"{e}l}(2006)}]{vilmin06EPJE}
\bibinfo{author}{\bibfnamefont{T.}~\bibnamefont{Vilmin}} \bibnamefont{and}
\bibinfo{author}{\bibfnamefont{E.}~{\bibnamefont{Rapha\"{e}l}}}, \bibinfo{journal}{Eur. Phys. J. E} \textbf{\bibinfo{volume}{21}}, \bibinfo{pages}{161} (\bibinfo{year}{2006}).

\bibitem[{\citenamefont{B\"{a}umchen et~al.}(2009)\citenamefont{B\"{a}umchen, Fetzer, and Jacobs}}]{baumchen09PRL}
\bibinfo{author}{\bibfnamefont{O.}~\bibnamefont{B\"{a}umchen}},
\bibinfo{author}{\bibfnamefont{R.}~\bibnamefont{Fetzer}}, \bibnamefont{and}
\bibinfo{author}{\bibfnamefont{K.}~\bibnamefont{Jacobs}},
\bibinfo{journal}{Phys. Rev. Lett.} \textbf{\bibinfo{volume}{103}},
\bibinfo{pages}{247801} (\bibinfo{year}{2009}).
  
\bibitem[{\citenamefont{Snoeijer and Eggers}(2010)}]{snoeijer10PRE}
\bibinfo{author}{\bibfnamefont{J. H.}~\bibnamefont{Snoeijer}} \bibnamefont{and}
\bibinfo{author}{\bibfnamefont{J.}~\bibnamefont{Eggers}},
\bibinfo{journal}{Phys. Rev. E} \textbf{\bibinfo{volume}{82}},
\bibinfo{pages}{056314} (\bibinfo{year}{2010}).
  
\bibitem[{\citenamefont{B\"{a}umchen et~al.}(2012)\citenamefont{B\"{a}umchen, Fetzer, Klos, Lessel, Marquant, H\"ahl, and Jacobs}}]{baumchen2012JCP}
\bibinfo{author}{\bibfnamefont{O.}~\bibnamefont{B\"{a}umchen}},
\bibinfo{author}{\bibfnamefont{R.}~\bibnamefont{Fetzer}},
\bibinfo{author}{\bibfnamefont{M.}~\bibnamefont{Klos}},
\bibinfo{author}{\bibfnamefont{M.}~\bibnamefont{Lessel}},
\bibinfo{author}{\bibfnamefont{L.}~\bibnamefont{Marquant}},
\bibinfo{author}{\bibfnamefont{H.}~\bibnamefont{H\"ahl}}, \bibnamefont{and}
\bibinfo{author}{\bibfnamefont{K.}~\bibnamefont{Jacobs}},
\bibinfo{journal}{J. Phys.-Condens. Mat.}
\textbf{\bibinfo{volume}{24}}, \bibinfo{pages}{325102}
(\bibinfo{year}{2012}).

\bibitem[{\citenamefont{Yang et~al.}(2010{\natexlab{b}})\citenamefont{Yang,Fujii, Lee, Lam, and Tsui}}]{Tsui10}
\bibinfo{author}{\bibfnamefont{Z.} \bibnamefont{Yang}},
\bibinfo{author}{\bibfnamefont{Y.}~\bibnamefont{Fujii}},  
\bibinfo{author}{\bibfnamefont{F. K.}~\bibnamefont{Lee}},
\bibinfo{author}{\bibfnamefont{C.-H.}~\bibnamefont{Lam}},
\bibnamefont{and}
\bibinfo{author}{\bibfnamefont{O. K. C.}~\bibnamefont{Tsui}},
\bibinfo{journal}{Science}
\textbf{\bibinfo{volume}{328}}, \bibinfo{pages}{1676} (\bibinfo{year}{2010}).

\bibitem[{\citenamefont{Barbero and Steiner}(2009)}]{barbero09PRL}
\bibinfo{author}{\bibfnamefont{D. R.}~\bibnamefont{Barbero}} \bibnamefont{and}
\bibinfo{author}{\bibfnamefont{U.}~\bibnamefont{Steiner}},
\bibinfo{journal}{Phys. Rev. Lett.} \textbf{\bibinfo{volume}{102}},
\bibinfo{pages}{248303} (\bibinfo{year}{2009}).

\bibitem{closa} F. Closa, F. Ziebert, and E. Rapha\"el, Phys. Rev. E \textbf{83}, 051603 (2011).

\bibitem[{\citenamefont{Rognin et~al.}(2011)\citenamefont{Rognin, Landis, and Davoust}}]{rognin11PRE}
\bibinfo{author}{\bibfnamefont{E.}~\bibnamefont{Rognin}},
\bibinfo{author}{\bibfnamefont{S.}~\bibnamefont{Landis}}, \bibnamefont{and}
\bibinfo{author}{\bibfnamefont{L.}~\bibnamefont{Davoust}},
\bibinfo{journal}{Phys. Rev. E} \textbf{\bibinfo{volume}{84}},
\bibinfo{pages}{041805} (\bibinfo{year}{2011}).

\bibitem[{\citenamefont{Rognin et~al.}(2012)\citenamefont{Rognin, Landis, and Davoust}}]{rognin12JVS}
\bibinfo{author}{\bibfnamefont{E.}~\bibnamefont{Rognin}},
\bibinfo{author}{\bibfnamefont{S.}~\bibnamefont{Landis}}, \bibnamefont{and}
\bibinfo{author}{\bibfnamefont{L.}~\bibnamefont{Davoust}},
\bibinfo{journal}{J. Vac. Sci. Technol., B}
\textbf{\bibinfo{volume}{30}}, \bibinfo{pages}{011602} (\bibinfo{year}{2012}).

\bibitem[{\citenamefont{McGraw et~al.}(2011)\citenamefont{McGraw, Jago, and Dalnoki-Veress}}]{jdmlev1}
\bibinfo{author}{\bibfnamefont{J. D.}~\bibnamefont{McGraw}},
\bibinfo{author}{\bibfnamefont{N. M.}~\bibnamefont{Jago}}, \bibnamefont{and}
\bibinfo{author}{\bibfnamefont{K.}~\bibnamefont{Dalnoki-Veress}},
\bibinfo{journal}{Soft Matter} \textbf{\bibinfo{volume}{7}},
\bibinfo{pages}{7832} (\bibinfo{year}{2011}).   
  
\bibitem[{\citenamefont{McGraw et~al.}(2012{\natexlab{b}})\citenamefont{McGraw, Salez, B\"{a}umchen, Rapha\"el, and Dalnoki-Veress}}]{McGraw12}
\bibinfo{author}{\bibfnamefont{J.~D.} \bibnamefont{McGraw}},
\bibinfo{author}{\bibfnamefont{T.}~\bibnamefont{Salez}},  
\bibinfo{author}{\bibfnamefont{O.}~\bibnamefont{B\"{a}umchen}},
\bibinfo{author}{\bibfnamefont{E.}~\bibnamefont{Rapha\"el}},
\bibnamefont{and}
\bibinfo{author}{\bibfnamefont{K.}~\bibnamefont{Dalnoki-Veress}},
\bibinfo{journal}{Phys. Rev. Lett.}
\textbf{\bibinfo{volume}{109}}, \bibinfo{pages}{128303} (\bibinfo{year}{2012}).
 
\bibitem[{\citenamefont{Salez et~al.}(2012{\natexlab{b}})\citenamefont{Salez, McGraw, B\"{a}umchen, Dalnoki-Veress, and Rapha\"el}}]{Salez12b}
\bibinfo{author}{\bibfnamefont{T.}~\bibnamefont{Salez}},
\bibinfo{author}{\bibfnamefont{J.~D.} \bibnamefont{McGraw}},
\bibinfo{author}{\bibfnamefont{O.}~\bibnamefont{B\"{a}umchen}},
\bibinfo{author}{\bibfnamefont{K.}~\bibnamefont{Dalnoki-Veress}},
\bibnamefont{and}
\bibinfo{author}{\bibfnamefont{E.}~\bibnamefont{Rapha\"el}},
\bibinfo{journal}{Phys. Fluids}
\textbf{\bibinfo{volume}{24}}, \bibinfo{pages}{102111} (\bibinfo{year}{2012}).

\bibitem[{\citenamefont{Salez et~al.}(2012{\natexlab{a}})\citenamefont{Salez, McGraw, Cormier, B\"{a}umchen, Dalnoki-Veress, and Rapha\"el}}]{salez12a}
\bibinfo{author}{\bibfnamefont{T.}~\bibnamefont{Salez}},
\bibinfo{author}{\bibfnamefont{J.~D.} \bibnamefont{McGraw}},
\bibinfo{author}{\bibfnamefont{S.~L.} \bibnamefont{Cormier}},
\bibinfo{author}{\bibfnamefont{O.}~\bibnamefont{B\"{a}umchen}},
\bibinfo{author}{\bibfnamefont{K.}~\bibnamefont{Dalnoki-Veress}},
\bibnamefont{and}
\bibinfo{author}{\bibfnamefont{E.}~\bibnamefont{Rapha\"el}},
\bibinfo{journal}{Eur. Phys. J. E}
\textbf{\bibinfo{volume}{35}}, \bibinfo{pages}{114} (\bibinfo{year}{2012}).

\bibitem[{\citenamefont{McGraw et~al.}(2013{\natexlab{b}})\citenamefont{McGraw, Salez, B\"{a}umchen, Rapha\"el, and Dalnoki-Veress}}]{McGraw13}
\bibinfo{author}{\bibfnamefont{J.~D.} \bibnamefont{McGraw}},
\bibinfo{author}{\bibfnamefont{T.}~\bibnamefont{Salez}},  
\bibinfo{author}{\bibfnamefont{O.}~\bibnamefont{B\"{a}umchen}},
\bibinfo{author}{\bibfnamefont{E.}~\bibnamefont{Rapha\"el}},
\bibnamefont{and}
\bibinfo{author}{\bibfnamefont{K.}~\bibnamefont{Dalnoki-Veress}},
\bibinfo{journal}{Soft Matter}
\textbf{\bibinfo{volume}{9}}, \bibinfo{pages}{8297} (\bibinfo{year}{2013}).

\bibitem[{\citenamefont{Barenblatt}(2003)}]{baren03TXT}
\bibinfo{author}{\bibfnamefont{G. I.}~\bibnamefont{Barenblatt}} 
\emph{\bibinfo{title}{Scaling, Self-Similarity, and Intermediate Asymptotics}} (\bibinfo{publisher}{Cambridge University Press, Cambridge}, \bibinfo{year}{1996}).

\bibitem{Reynolds}O. Reynolds, Phil. Trans. Roy. Soc. London, \textbf{186} 123 (1895).

\bibitem{Taylor1} G. I. Taylor, Proc. Roy. Soc., \textbf{201} 159 (1950).

\bibitem{Taylor2} G. I. Taylor, Proc. Roy. Soc., \textbf{201} 175 (1950).

\bibitem[{\citenamefont{Benzaquen et~al.}(2013)\citenamefont{Benzaquen, Salez, and Rapha\"el}}]{Benzaquen13}
\bibinfo{author}{\bibfnamefont{M.} \bibnamefont{Benzaquen}},
\bibinfo{author}{\bibfnamefont{T.}~\bibnamefont{Salez}},  
\bibnamefont{and}
\bibinfo{author}{\bibfnamefont{E.}~\bibnamefont{Rapha\"el}},
\bibinfo{journal}{Eur. Phys. J. E}
\textbf{\bibinfo{volume}{36}}, \bibinfo{pages}{82} (\bibinfo{year}{2013}).

\bibitem{note1}We have previously verified that for these experiments there is no difference in annealing in an ambient or inert environment.

\bibitem[{\citenamefont{de~{G}ennes et~al.}(2003)\citenamefont{de~{G}ennes, Brochard-Wyart, and Qu\'{e}r\'{e}}}]{degennes03TXT}
\bibinfo{author}{\bibfnamefont{P.-G.}~\bibnamefont{de~{G}ennes}},
\bibinfo{author}{\bibfnamefont{F.}~\bibnamefont{Brochard-Wyart}},
\bibnamefont{and}
\bibinfo{author}{\bibfnamefont{D.}~\bibnamefont{Qu\'{e}r\'{e}}},
\emph{\bibinfo{title}{Capillarity and {W}etting {P}henomena: {D}rops, {B}ubbles, {P}earls, {W}aves}} (\bibinfo{publisher}{Springer}, \bibinfo{year}{2003}).

\bibitem[{\citenamefont{Huppert}(1982)}]{refBcite}
\bibinfo{author}{\bibfnamefont{H.}~\bibnamefont{Huppert}},
\bibinfo{journal}{J. Fluid Mech.} \textbf{\bibinfo{volume}{121}},
\bibinfo{pages}{43} (\bibinfo{year}{1982}).

\bibitem[{\citenamefont{Seemann et~al.}(2001{\natexlab{b}})\citenamefont{Seemann, Herminghaus, and Jacobs}}]{seeman2001PRL}
\bibinfo{author}{\bibfnamefont{R.}~\bibnamefont{Seemann}},
\bibinfo{author}{\bibfnamefont{S.}~\bibnamefont{Herminghaus}},
\bibnamefont{and} \bibinfo{author}{\bibfnamefont{K.}~\bibnamefont{Jacobs}},
\bibinfo{journal}{Phys. Rev. Lett.} \textbf{\bibinfo{volume}{86}},
\bibinfo{pages}{5534} (\bibinfo{year}{2001}{\natexlab{b}}).

\bibitem[{\citenamefont{Bach et~al.}(2003)\citenamefont{Bach, Almdal, Rasmussen, and Hassager}}]{bach03MAC}
\bibinfo{author}{\bibfnamefont{A.}~\bibnamefont{Bach}},
\bibinfo{author}{\bibfnamefont{K.}~\bibnamefont{Almdal}},
\bibinfo{author}{\bibfnamefont{H.}~\bibnamefont{Rasmussen}},
\bibnamefont{and} \bibinfo{author}{\bibfnamefont{O.}~\bibnamefont{Hassager}},
\bibinfo{journal}{Macromolecules} \textbf{\bibinfo{volume}{36}},
\bibinfo{pages}{5174} (\bibinfo{year}{2003}).

\bibitem[{\citenamefont{Stillwagon and Larson}(1988)}]{stillwagon88JAP}
\bibinfo{author}{\bibfnamefont{L. G.}~\bibnamefont{Stillwagon}} \bibnamefont{and}
\bibinfo{author}{\bibfnamefont{R.}~\bibnamefont{Larson}},
\bibinfo{journal}{Journal of Applied Physics} \textbf{\bibinfo{volume}{63}},
\bibinfo{pages}{5251} (\bibinfo{year}{1988}).

\bibitem[{\citenamefont{Oron et~al.}(1997)\citenamefont{Oron, Davis, and Bankoff}}]{oron97RMP}
\bibinfo{author}{\bibfnamefont{A.}~\bibnamefont{Oron}},
\bibinfo{author}{\bibfnamefont{S.}~\bibnamefont{Davis}}, \bibnamefont{and}
\bibinfo{author}{\bibfnamefont{S.}~\bibnamefont{Bankoff}},
\bibinfo{journal}{Rev. Mod. Phys.} \textbf{\bibinfo{volume}{69}},
\bibinfo{pages}{931} (\bibinfo{year}{1997}).

\bibitem[{\citenamefont{Craster and Matar}(2009)}]{craster09RMP}
\bibinfo{author}{\bibfnamefont{R.}~\bibnamefont{Craster}} \bibnamefont{and}
 \bibinfo{author}{\bibfnamefont{O.}~\bibnamefont{Matar}},
\bibinfo{journal}{Rev. Mod. Phys.} \textbf{\bibinfo{volume}{81}},
\bibinfo{pages}{1131} (\bibinfo{year}{2009}).

\bibitem{blossey} R. Blossey, \emph{Thin Liquid Films}. ISBN 9789400744547, Springer, Dordrecht, (2012).

\bibitem[{\citenamefont{Bertozzi}(1998)}]{bertozziAMS98}
\bibinfo{author}{\bibfnamefont{A.}~\bibnamefont{Bertozzi}},
\bibinfo{journal}{Notices Amer. Math. Soc.} \textbf{\bibinfo{volume}{45}},
\bibinfo{pages}{689} (\bibinfo{year}{1998}).

\bibitem[{\citenamefont{Wu}(1999)}]{wuPpoly}
\bibinfo{author}{\bibfnamefont{S.}~\bibnamefont{Wu}},
\emph{\bibinfo{title}{Polymer Handbook}}, vol.~\bibinfo{volume}{4}
\bibinfo{publisher}{Wiley-Interscience}, (\bibinfo{year}{1999}).

\end{thebibliography}
\end{document}